\newcounter{myctr}
\def\myitem{\refstepcounter{myctr}\bibfont\noindent\ifnum\themyctr>9\else\phantom{0}\fi\hangindent17pt\themyctr.\enskip}

\documentclass{article}

\usepackage{amsmath}
\usepackage{amsfonts}
\usepackage{graphicx}
\usepackage{fullpage}
\newcommand{\ket}[1]{\ensuremath{|#1\rangle}}
\newcommand{\bra}[1]{\ensuremath{\langle#1|}}

\newcommand{\Id}{\mathbb{I}}

\begin{document}
\markboth{Piotr Gawron, Jaros{\l}aw Miszczak, Jan S{\l}adkowski}
{Noise Effects in Quantum Magic Squares Game}

\title{Noise Effects in Quantum Magic Squares Game}

\author{PIOTR GAWRON\\
JAROS{\L}AW MISZCZAK\\
Institute of Theoretical and Applied Informatics,\\
Polish Academy of Sciences,
ul. Ba{\l}tycka 5\\
Gliwice, 44-100, Poland\\
\{gawron,miszczak\}@iitis.gliwice.pl
\\[1em]
JAN S{\L}ADKOWSKI\\
Institute of Physics, University of Silesia, ul. Bankowa 14\\
Katowice, 40-007, Poland\\
jan.sladkowski@us.edu.pl
}
\maketitle

\begin{abstract}
In the article we analyse how noisiness of quantum channels
can influence the magic squares quantum pseudo-telepathy game.
We show that the probability of success can be used to determine
characteristics of quantum channels. Therefore the game deserves
more careful study aiming at its implementation. 
\end{abstract} 

\section{Introduction}
Quantum game theory, the
subclass of game theory that involves quantum phenomena\cite{EWL,PS}, 
lies at the crossroads of physics, quantum
information processing, computer and natural sciences. Thanks to
entanglement, quantum players\footnote{By a quantum player we
understand a player that, at least in theory, can explore and make
profits from the quantum phenomena in situations of conflict,
rivalry etc.} can sometimes accomplish tasks that are impossible
for their classical counterparts. In this paper we present the
detailed analysis of one of the pseudo-telepathy games\cite{brassard04pseudo,brass2}. 
These games provide simple, yet nontrivial, examples of quantum
games that can be used to show the effects of quantum
non-local correlations. Roughly speaking, a~game belongs to the
pseudo-telepathy class if it admits no winning strategy for
classical players, but it admits a winning strategy provided the
players share the~sufficient amount of entanglement. This
phenomenon is called pseudo-telepathy, because it would appear as
magical to a classical player, yet it has quantum theoretical
explanation. 

Our main goal is to study the
connection between errors in quantum channels and the probability
of winning in magic squares game. The magic square game is
selected because of its unique features: it is easy to show that
there is no classical winning strategy, simple enough for a
layman to follow its course and feasible.

The paper is organized as follows. We will begin by presenting the 
magic squares pseudo-telepathy game. The we will introduce tools we 
are using to analyse noise in quantum systems.
Then we will attempt to answer the
following question: {\it What happens if a quantum game is played
in non-perfect conditions because of the influence of quantum
noise?} The results showing the connection between the noise level
and the probability of winning will be given in Section 4.
Finally we will point out some issues that yet should be addressed.

\section{Magic square game}
The magic square is a~$3\times 3$ matrix filled
with numbers 0 or 1 so that the sum of entries in each row is
even and the sum of entries in each column is odd. Although such a~matrix 
cannot exist\footnote{Therefore the adjective {\it magic}
is used.} one can consider the following game.

There are two players: Alice and Bob.
Alice is given the number of the row, Bob is given the number of the column.
Alice has to give the entries for a~row and Bob has to give entries for a~column
so that the parity conditions are met. In addition, the
intersection of the row and the column must agree. Alice and Bob can prepare a strategy 
but they are not allowed to communicate during the game.

There exists a (classical)
strategy that leads to winning probability of $\frac{8}{9}$. 
If parties are allowed to share a~quantum  state they can achieve
probability $1$. 

In the quantum version of this game\cite{Mer}
Alice and Bob are allowed to share an entangled quantum state.

The winning strategy is following.
Alice and Bob share entangled state
\begin{equation}
\ket{\Psi}=\frac{1}{2}(\ket{0011}-\ket{1100}-\ket{0110}+\ket{1001}). 
\end{equation}
Depending on the input (\emph{i.e.} the specific row and column to be filled in)
Alice and Bob apply unitary operators $A_i\otimes
\Id$ and $\Id\otimes B_j$, respectively, 
{
\begin{eqnarray}
A_1=\frac{1}{\sqrt{2}}
\left[
\begin{array}{cccc}
i & 0 & 0 & 1\cr
0 &-i & 1 & 0\cr
0 & i & 1 & 0\cr
1 & 0 & 0 & i
\end{array}
\right]
&
A_2=
\frac{1}{2}
\left[
\begin{array}{cccc}
i & 1 & 1 & i\cr
-i & 1 & -1 & i\cr
i & 1 & -1 & -i\cr
-i & 1 & 1 & -i
\end{array}
\right]
&
A_3=
\frac{1}{2}
\left[
\begin{array}{cccc}
-1 & -1 & -1 & 1\cr
1 & 1 & -1 & 1\cr
1 & -1 & 1 & 1\cr
1 & -1 & -1 & -1
\end{array}
\right]
\\
B_1=
\frac{1}{2}
\left[
\begin{array}{cccc}
i & -i & 1 & 1\cr
-i & -i & 1 & -1\cr
1 & 1 & -i & i\cr
-i & i & 1 & 1
\end{array}
\right]
&
B_2=
\frac{1}{2}
\left[
\begin{array}{cccc}
-1 & i & 1 & i\cr
1 & i & 1 & -i\cr
1 & -i & 1 & i\cr
-1 & -i & 1 & -i
\end{array}
\right]
&
B_3=
\frac{1}{\sqrt{2}}
\left[
\begin{array}{cccc}
1 & 0 & 0 & 1\cr
-1 & 0 & 0 & 1\cr
0 & 1 & 1 & 0\cr
0 & 1 & -1 & 0
\end{array}
\right]
\end{eqnarray}
}
where $i$ and $j$ denote the corresponding inputs.

The final state is used to determine two bits of
each answer. The remaining bits can be found by applying parity
conditions.

\section{Quantum noise}

A interesting question arises: what happens if a~quantum
game is played in non-perfect (real-world) conditions because of
the presence of quantum noise.\cite{CAKK,CKO}

In the most general case quantum evolution is described by superoperator 
$\Phi$, which can be expressed using Kraus representation\cite{nc}:
\begin{equation}
\Phi(\rho)=\sum_k E_k \rho {E_k}^\dagger,
\end{equation}
where $\sum_k {E_k}^\dagger E_k=\Id$.

In following we will consider typical quantum channels, namely
\begin{itemize}
 \item depolarizing channel:
$\left\{
\sqrt{1-\frac{3\alpha}{4}}\Id
,
\sqrt{\frac{\alpha}{4}}\sigma_x
,
\sqrt{\frac{\alpha}{4}}\sigma_y,
\sqrt{\frac{\alpha}{4}}\sigma_z
\right\}$,
\item amplitude damping:
$\left\{
\left[
\begin{array}{cc}
1 & 0 \\
0 & \sqrt{1-\alpha}
\end{array}
\right]
,
\left[
\begin{array}{cc}
0 & \sqrt{\alpha} \\
0 & 0
\end{array}
\right]
\right\}$,

\item phase flip, bit flip and bit-phase flip with Kraus operators\\ $\left\{
\sqrt{1-\alpha}\Id,
\sqrt{\alpha}\sigma_z
\right\}
$,
$
\left\{
\sqrt{1-\alpha}\Id,
\sqrt{\alpha}\sigma_x
\right\}
$
and 
$\left\{
\sqrt{1-\alpha}\Id
,
\sqrt{\alpha}\sigma_y
\right\}$ 
respectively.
\end{itemize}
Real parameter $\alpha\in [0,1]$ represents here the amount of noise in the channel and 
$\sigma_x,\sigma_y,\sigma_z$ are Pauli matrices.

In our scheme, the Kraus operators are of the dimension $2^4$. 
They are constructed from one-qubit operators $e_k$ by taking their tensor product 
over all $n^4$ combinations of $\pi(i)$ indices
\begin{equation}
E_k=\bigotimes_\pi e_{\pi(i)},
\end{equation}
where $n$ is the number of Kraus operator for a single qubit channel.

\subsection{The comparison of channels}

Although one can assign physical meaning to the parameter $\alpha$, this meaning can be 
different for different channels. Therefore we are using channel fidelity\cite{GLN} to 
compare quantum channels.

We define channel fidelity as:
\begin{equation}
 \Delta(\Phi)=F(J(\Phi), J(\Id)),
\end{equation}
where $J$ is Jamio{\l}kowski isomorphism and $F$ is the fidelity defined as 
$F(\rho_1,\rho_2)={tr(\sqrt{\sqrt{\rho_1}\rho_2\sqrt{\rho_1}})}^2$.

\section{Results}
In this section we are analysing the
influence of the noise on success probability and fidelity of
non-perfect (mixed) quantum states in the case when the noise
operator is applied before the game gates.
\subsection{Calculations}
The final state of this scheme is $\rho_f=(A_i\otimes B_j)\,\Phi_\alpha(\ket{\Psi}\bra{\Psi})\,(A_i^\dagger\otimes B_j^\dagger)$, where $\Phi_\alpha$ 
is the superoperator realizing quantum channel parametrized by real number $\alpha$. Probability $P_{i,j}(\alpha)$ is computed as the probability 
of measuring $\rho_f$ in the state indicating success
\begin{equation}
P_{i,j}(\alpha)=tr\left(\rho_f \sum_i{\ket{\xi_i}\bra{\xi_i}}\right),
\end{equation}
where $\ket{\xi_i}$ are the states that imply success.

\subsection{Success probability}
\begin{figure}
  \includegraphics[width=.49\textwidth]{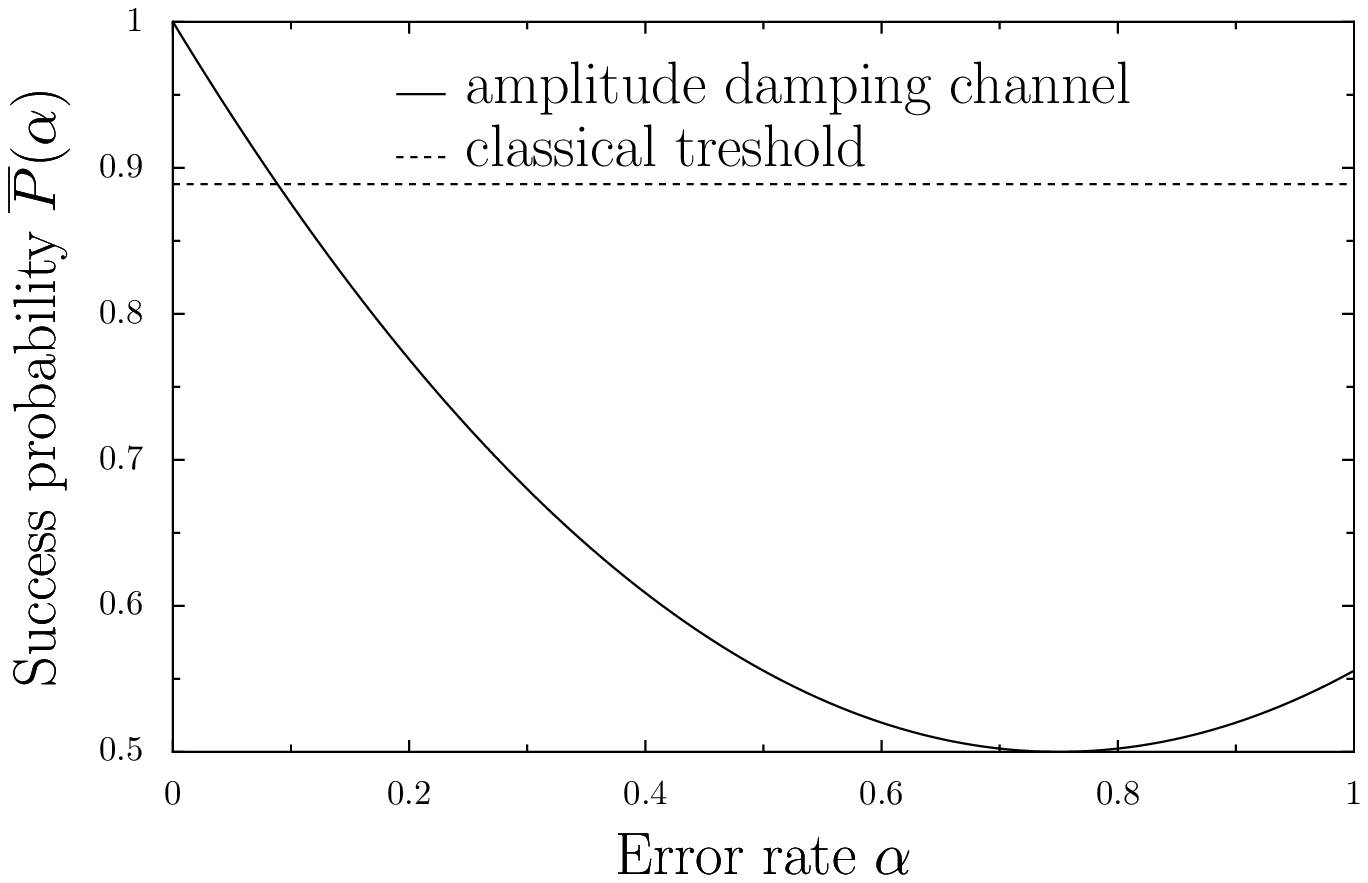}
  \includegraphics[width=.49\textwidth]{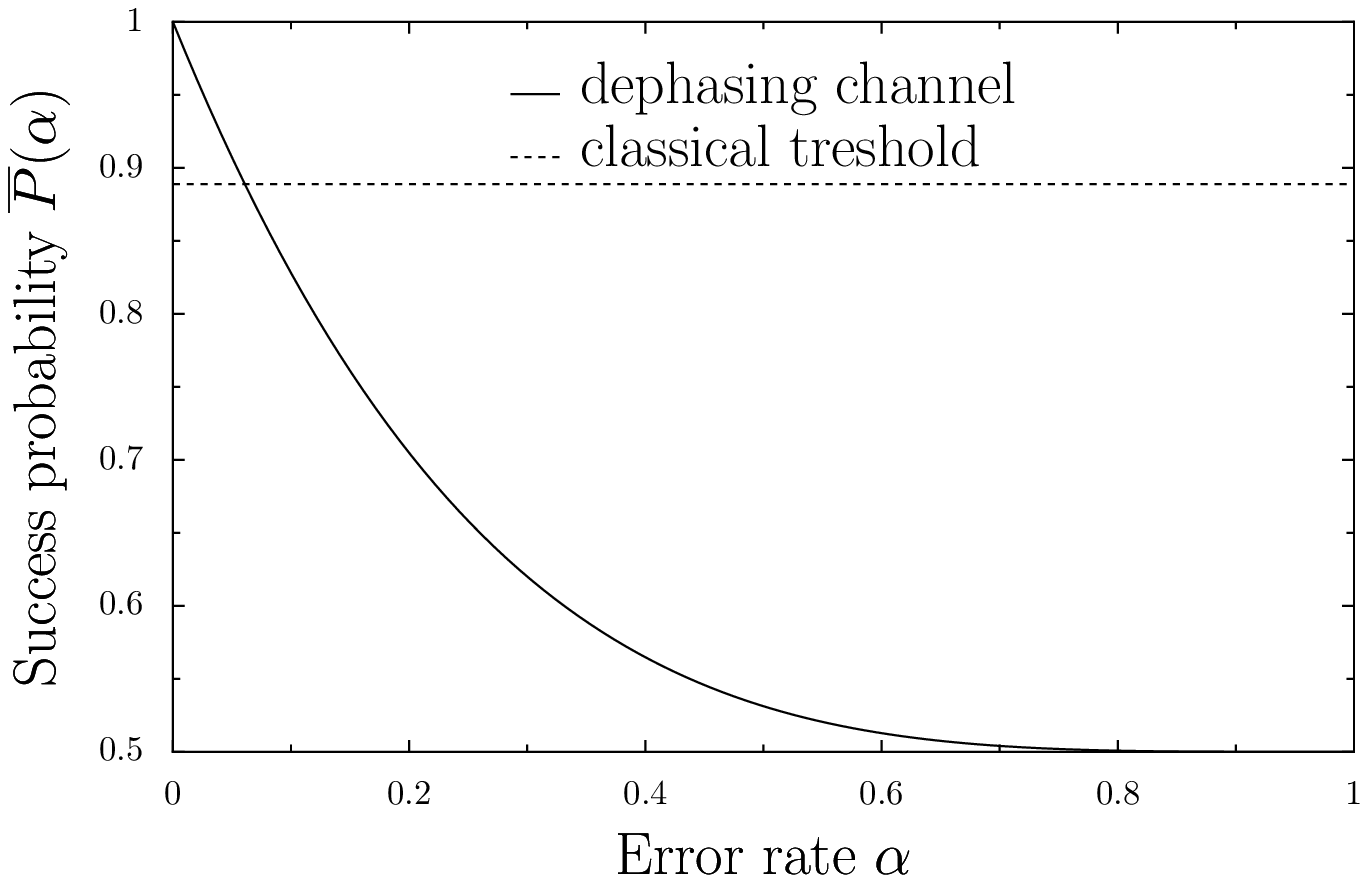}
  \includegraphics[width=.49\textwidth]{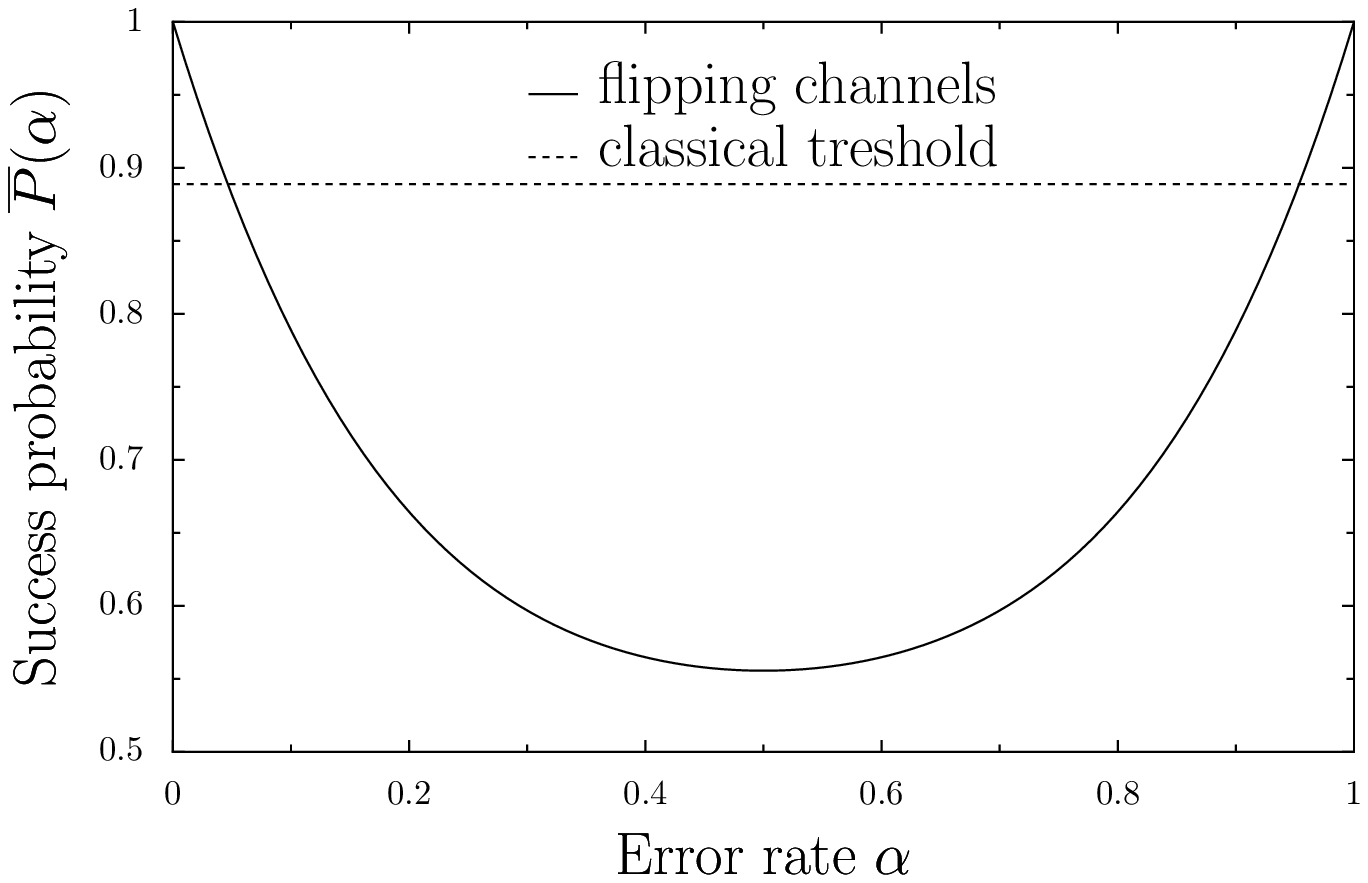}
  \includegraphics[width=.49\textwidth]{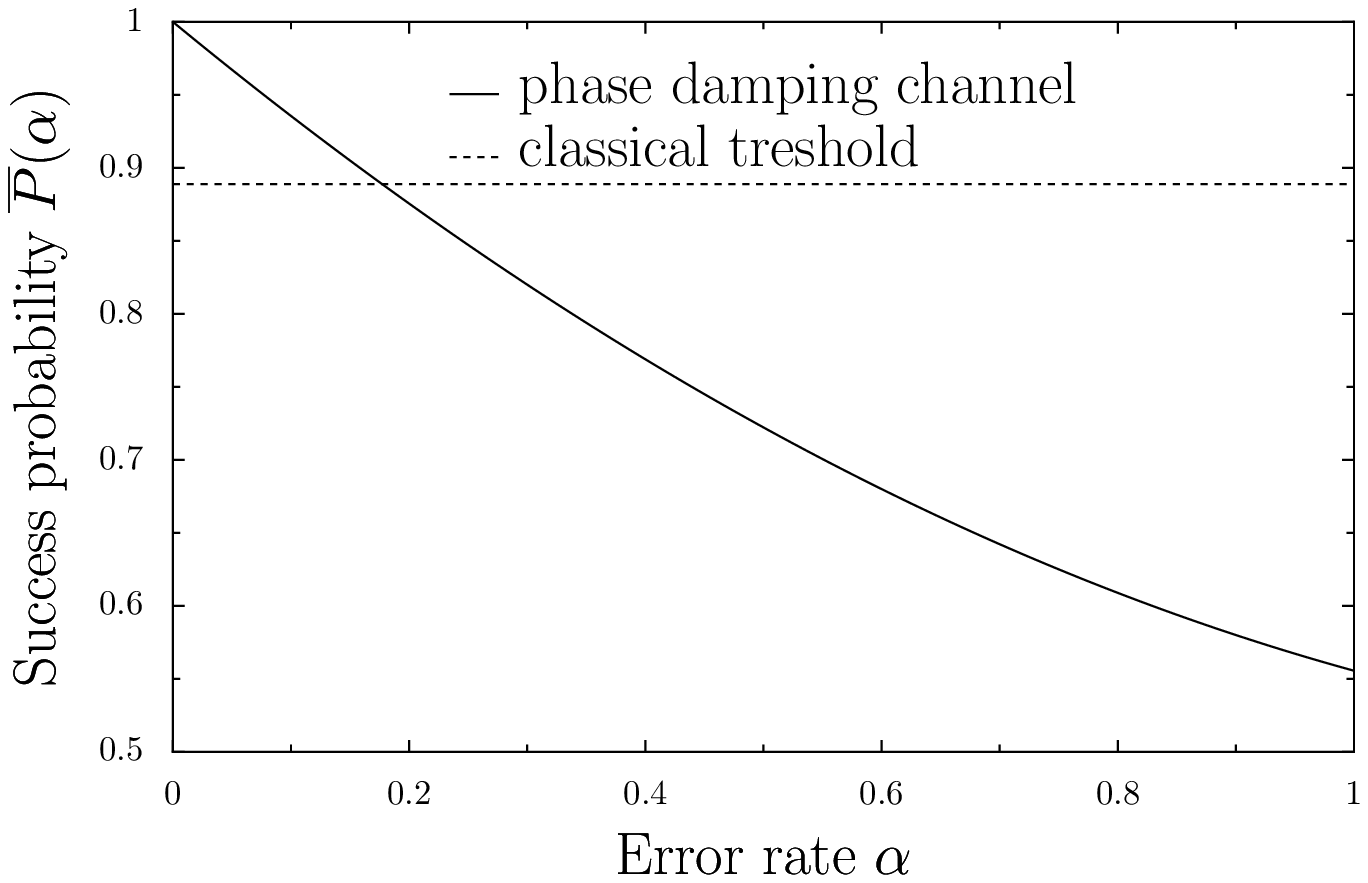}
  \caption{Dephasing and damping channels cause monotonic decrease of mean success probability in the function of noise amount. Amplitude damping channel causes the success function to attain minimal probability of success for error rate $\frac{3}{4}$. Flipping channels give symmetrical functions with minimum for error rate $\frac{1}{2}$.}
\label{fig:prob2}
\end{figure}

We compute success probability $P_{i,j}(\alpha)$ for different inputs ($i,j\in\{1,2,3\}$)
and different quantum channels.
Our calculations show that mean probability of~success,
$\overline{P}(\alpha)=\sum_{i,j\in\{1,2,3\}} P_{i,j}(\alpha)$, heavily
depends on the noise level $\alpha$. 
The game results for each combination of gates $A_i, B_j$ for depolarizing,
amplitude damping, phase damping, phase, bit and bit-phase flip channels are listed in
Fig.~\ref{fig:prob1}. Fig.~\ref{fig:prob2} presents mean success probability 
$\overline{P}(\alpha)$ as the function of error rate.

\begin{figure}
\begin{center}
\begin{tabular}{|lrcl|}
\hline
\multicolumn{4}{|c|}{Depolarizing channel:}\\
 $\{(i,j)| i,j=1,2,3\}$ & $P_{i,j}(\alpha)$&=&$\frac{1}{2}\alpha^4-2\,\alpha^3+3\,\alpha^2-2\,\alpha+1$.\\
\hline
\hline
\multicolumn{4}{|c|}{Amplitude damping channel:}\\
$(i,j)\in\{(1,1), (1,2), (2,3), (3,3)\}$ & $P_{i,j}(\alpha)$&=&$\frac{1}{2}\alpha^2-\alpha+1$\\
$(i,j)\in\{(1,3)\}$ & $P_{i,j}(\alpha)$&=&$2\alpha^2-2\,\alpha+1$\\
$(i,j)\in\{(2,1), (2,2), (3,1), (3,2)\}$ & $P_{i,j}(\alpha)$&=&$\alpha^2-\frac{3}{2}\,\alpha+1$\\
\hline
\hline
\multicolumn{4}{|c|}{Phase damping channel:}\\
$(i,j)\in\{(1,1), (1,2), (2,3), (3,3)\}$ & $P_{i,j}(\alpha)$&=&$\frac{1}{2}\alpha^2-\alpha+1$\\
$(i,j)\in\{(1,3)\}$ & $P_{i,j}(\alpha)$&=&$1$\\
$(i,j)\in\{(2,1), (2,2), (3,1), (3,2)\}$ & $P_{i,j}(\alpha)$&=&$-\frac{1}{2}\,\alpha+1$\\
\hline
\hline
\multicolumn{4}{|c|}{Phase flip:}\\
$(i,j)\in\{(1,1), (1,2), (2,3), (3,3)\}$ & $P_{i,j}(\alpha)$&=&$8\,\alpha^4-16\,\alpha^3+12\,\alpha^2-4\,\alpha+1$ \\
$(i,j)\in\{(1,3)\}$ & $P_{i,j}(\alpha)$&=&$1$\\
$(i,j)\in\{(2,1), (2,2), (3,1), (3,2)\}$ & $P_{i,j}(\alpha)$&=&$2\,\alpha^2-2\,\alpha+1$\\
\hline
\hline
\multicolumn{4}{|c|}{Bit flip:}\\
$(i,j)\in\{(1,1), (1,2), (3,1), (3,2)\}$ & $P_{i,j}(\alpha)$&=&$2\,\alpha^2-2\,\alpha+1$ \\
$(i,j)\in\{(2,3)\}$ & $P_{i,j}(\alpha)$&=&$1$\\
$(i,j)\in\{(1,3), (2,1), (2,2), (3,3)\}$ & $P_{i,j}(\alpha)$&=&$8\,\alpha^4-16\,\alpha^3+12\,\alpha^2-4\,\alpha+1$\\
\hline
\hline
\multicolumn{4}{|c|}{Bit-phase flip:}\\
$(i,j)\in\{(1,1), (1,2), (2,1), (2,2)\}$ & $P_{i,j}(\alpha)$&=&$2\,\alpha^2-2\,\alpha+1$ \\
$(i,j)\in\{(3,3)\}$ & $P_{i,j}(\alpha)$&=&$1$\\
$(i,j)\in\{(1,3), (2,3), (3,1), (3,2)\}$ & $P_{i,j}(\alpha)$&=&$8\,\alpha^4-16\,\alpha^3+12\,\alpha^2-4\,\alpha+1$ \\
\hline
\end{tabular}
\caption{Success probability for all combinations of magic squares game inputs for depolarizing, amplitude damping, phase damping channels, phase, bit and bit-phase flip channels}
\label{fig:prob1}
\end{center}
\end{figure}

In the case of depolarizing channel, the success probability as the function
of noise amount is the same for all the possible inputs. In the
case of~amplitude and phase damping channels, wan can observe three
different types of behaviour. These functions are non-increasing
for those channels. The bit, phase and bit-phase flip functions
reach their minima for $\alpha=1/2$ and are symmetrical. This
means that high error rates influence the game weakly. One can
easily see that in case of input $(1,3)$ the phase-flip channel
does not influence the probability of success. The same is true
for input $(2,3)$ and bit flip channel and also for input $(3,3)$
and bit-phase flip channel. Therefore it is possible to
distinguish those channels by looking at success probability of
magic-squares game. 

\begin{figure}[htpb!]
 \centering
 \includegraphics[width=.92\textwidth]{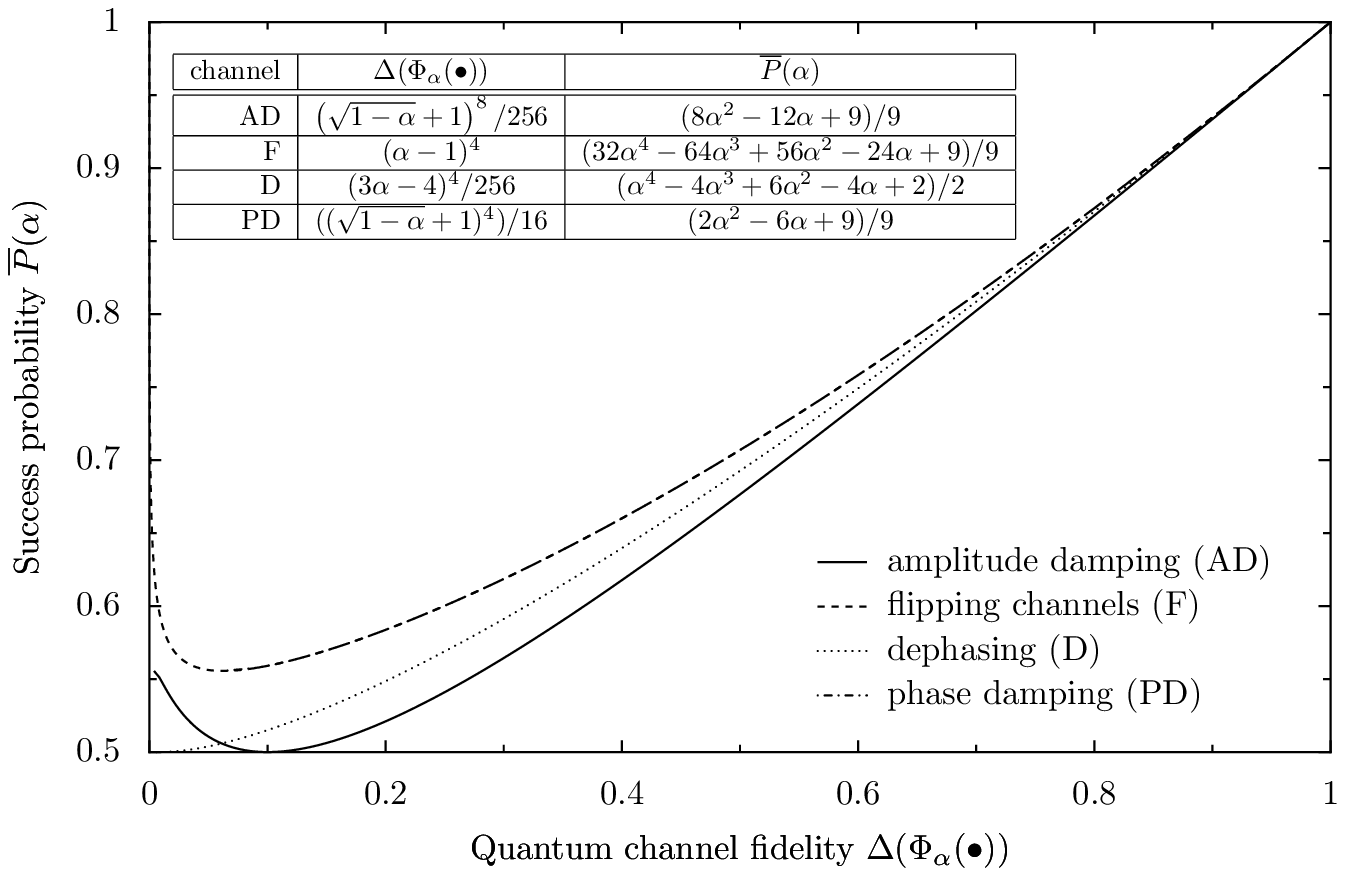}
 \caption{The influence of noise on success probability, in case of different quantum channels is compared by using the parametric plot of success probability $P_{i,j}(\alpha)$ versus quantum channel fidelity $\Delta(\Phi(\alpha))$ for $\alpha\in[0,1]$. Note that plots for flipping channels and phase damping channel overalp in range $\Delta(\Phi(\alpha))\in\left[\frac{1}{16},1\right]$.}
\label{fig:prob-fid-plot}
\end{figure}

The graphical representation of dependency between mean success probability and channel fidelity is presented in the form of parametric plot in Fig.~\ref{fig:prob-fid-plot}.

\section{Conclusions}
We have shown how the probability success in magic squares
pseudo-telepathy game is influenced by different quantum noisy
channels. The calculations show that, by controlling noise parameter and observing probabilities of success,
it is possible to distinguish some channels. Thus we have shown that
implementation of magic square game can provide the example 
of channel distinguishing procedure.

In case of all channels success probability drops,
with the increase of noise, below classical limit of $\frac{8}{9}$. 
Therefore the physical implementation of quantum magic squares
game requires high precision and can be a very difficult task. 

We have also shown that if channel fidelity is higher than $1/10$ 
the probability of success is almost linear. Therefore channel fidelity is good approximation of success 
probability for not very noisy channels.

\section*{Acknowledgements}
This research was supported in part by the Polish Ministry of Science  and Higher Education project No N519 012 31/1957.

\end{document}